\RequirePackage{lineno}
\documentclass[aps, prl,
twocolumn,superscrptaddress,showpacs]{revtex4-1}
\usepackage{t1enc}
\usepackage[english]{babel}
\usepackage{graphicx}
\usepackage{epsfig}
\usepackage{amssymb}
\usepackage{dcolumn}
\usepackage{bm}
\usepackage{color}
\usepackage{float}
\usepackage{natbib}

\newcommand{\NCCO}{\ensuremath{\mathrm{Nd}_{2-x}\mathrm{Ce}_{x}\mathrm{Cu}\mathrm{O}_{4}}}
\newcommand{\LSCO}{\ensuremath{\mathrm{La}_{2-x}\mathrm{Sr}_{x}\mathrm{Cu}\mathrm{O}_{4}}}

\newcommand{\dg}{\ensuremath{^\circ}}

\begin{document}

\title{The high-energy anomaly in  ARPES spectra of the cuprates---many body  or matrix element effect?}
\author{E. D. L. Rienks,$^1$  M. \"Arr\"al\"a,$^2$  M. Lindroos,$^2$    F. Roth,$^3$ W. Tabis,$^{4,5}$  G. Yu,$^4$ M. Greven,$^4$ and    J. Fink$^6$}
\affiliation{
$^1$ Helmholtz-Zentrum Berlin, Albert-Einstein-Strasse 15, D-12489 Berlin, Germany\\
$^2$ Department of Physics, Tampere University of Technology, PO Box 692, FIN-33101 Tampere, Finland\\
$^3$ Center for Free-Electron Laser Science / DESY, Notkestrasse 85, D-22607 Hamburg, Germany\\
$^4$ School of Physics and Astronomy, University of Minnesota, Minneapolis, Minnesota 55455, USA\\
$^5$AGH University of Science and Technology, Faculty of Physics and Applied Computer Science, 30-059 Krakow, Poland\\
$^6$ Leibniz-Institute for Solid State and Materials Research Dresden, P.O.Box 270116, D-01171 Dresden, Germany\\
}

\date{\today}

\begin{abstract}
We used  polarization-dependent angle-resolved  photoemission spectroscopy (ARPES)   to study the
high-energy anomaly (HEA) in the dispersion   of \NCCO, ($x=0.123$). We have found that at particular photon energies the anomalous, waterfalllike dispersion gives way to a broad, continuous band. This suggests that the HEA is a matrix element effect: it arises due to a  suppression of the intensity of the broadened quasi-particle band in a narrow momentum range. We confirm this interpretation experimentally, by showing that the HEA appears when the matrix element is suppressed deliberately by changing the light polarization. Calculations of the matrix element using atomic wave functions and simulation of the ARPES intensity with one-step model  calculations provide further proof for this scenario.  The possibility to detect the full quasi-particle dispersion further allows us to extract the high-energy self-energy function near the center  and at the edge  of the Brillouin zone.
\end{abstract}


\maketitle

One of the unique assets of angle-resolved photoemission spectroscopy (ARPES) is the ability to determine  the spectral function $A(\omega,\boldsymbol{k})$  in energy and momentum space. The finite width and deviation of the dispersion from that calculated in an independent particle model are  interpreted in the majority of cases in terms of many-body effects\,\cite{Damascelli2003}. In the cuprate high-$T_c$ superconductors, various kinks in the dispersion have been discovered which were analyzed in terms of a coupling of the charge carriers to  bosonic excitations possibly mediating high-$T_c$ superconductivity in these materials. Besides the  kinks in the low binding energy $(E_B)$ region ($E_B \leq $ 0.1 eV) at  $E_B=E_H\gtrapprox 0.3$ eV the band appears to bend sharply and seems to  proceed almost vertically towards the valence bands. This phenomenon has been termed "waterfall" or high-energy anomaly (HEA)\,\cite{Graf2007}.  The HEA  has been observed in undoped cuprates\,\cite{Ronning2005} as well as in  their hole-doped\,\cite{Pan2006,Graf2007,Meevasana2007,Chang2007a,Xie2007,Inosov2007a,Valla2007,Inosov2008,Meevasana2008,Zhang2008,Moritz2009}, and electron-doped derivatives\,\cite{Pan2006,Ikeda2009a,Moritz2009,Schmitt2011}. In the latter two systems  $E_H$ shows a $d$-wave  momentum dependence being larger along the nodal direction and smaller near the antinodal point opposite to the momentum dependence of the $d$-wave superconducting gap\,\cite{Chang2007a,Zhang2008,Ikeda2009a}. The values of $E_H$ exhibit a difference of $\approx$ 0.4 eV between hole doped and electron doped cuprates\,\cite{Pan2006}. This difference was interpreted  in terms of a shift of the chemical potential\,\cite{Ikeda2009a}. The experimental studies were accompanied by numerous theoretical papers\,\cite{Markiewicz2007a,Markiewicz2007,Byczuk2007,Macridin2007,Srivastava2007,Tan2007,Leigh2007,Zhu2008,Wrobel2008,Weber2008,Zemljic2008,Matho2010,Sakai2010,Markiewicz2010,Moritz2010, Weber2008,Katagiri2011}.

For the phenomenon of the HEA, a number of explanations have been suggested including Mott-Hubbard models with a transition from the coherent quasi-particle dispersion to the incoherent lower Hubbard band\,\cite{Meevasana2007,Byczuk2007,Manousakis2007,Tan2007,Zemljic2008}, a disintegration of the low-energy branch into a holon and spinon band due to a spin charge separation\,\cite{Graf2007}, a coupling to spin fluctuations\,\cite{Valla2007,Macridin2007,Markiewicz2007,Basak2009,Markiewicz2010}, a coupling to phonons\,\cite{Xie2007}, string excitations of spin-polarons\,\cite{Manousakis2007}, a bifurcation  of the quasi-particle band due to an excitation of a bosonic mode of charge 2e\,\cite{Leigh2007}, and a coupling to plasmons\,\cite{Markiewicz2007a}. These are all intrinsic  interpretations in terms of many-body interactions leading to a change of the spectral function. 

However, the spectral function can strictly only be inferred from ARPES with a detailed knowledge of the photoexcitation matrix element, since the measured photocurrent is given by\,\cite{Damascelli2003}: 
\begin{equation}\label{eq:photo}
I(\omega,\boldsymbol{k})\propto |M(\omega,\boldsymbol{k})|^2 A(\omega,\boldsymbol{k})
\end{equation} 
where  the matrix element 
\begin{equation}\label{eq:matrix}
 M(\omega,\boldsymbol{k}) = \left< f \right|\boldsymbol{er} \left| i \right>
\end{equation}
is determined by the final state $\left< f \right|$, the initial state $\left| i \right>$, and  the dipole operator $\bf{er}$ ($\bf{e}$ is the unit vector along the polarization direction of the photons).

Some ARPES studies  pointed out that extrinsic effects due to matrix element effects may explain the HEA\,\cite{Inosov2007a,Inosov2008,Zhang2008} since changes of the waterfalllike dispersion to a Y-shaped  dispersion have been observed upon photon energy variation or by changing the Brillouin zone (BZ). Thereupon a combination of extrinsic and intrinsic effects have been invoked to explain the ARPES results of cuprates at high energies\,\cite{Meevasana2008,Moritz2009,Basak2009}.

In this Letter we address the controversy  of the explanation of the HEA in terms of extrinsic or intrinsic effects. We present a polarization and photon energy  dependent ARPES study on the electron-doped cuprate~\cite{Armitage2010} \NCCO\  ($x=0.123$)  in several BZs. We find that the waterfalllike dispersion transforms into a normal, dispersive band at certain photon energies. In addition, a waterfalllike dispersion can be induced in the intact band by changing the polarization of the incoming photons. The results can be explained in terms of a wipe-out of the intensity of the broadened quasi-particle band in a particular momentum range. Thus  we give strong evidence that the HEA is not caused by intrinsic many-body effects but rather by extrinsic matrix element effects. Furthermore, the newfound ability to observe the dispersion in the entire BZ  allows us to determine  the mass renormalization of the quasi-particle band relative to density functional theory (DFT) calculations.
  
The  \NCCO\ ($x=0.123$) single crystal was grown in about 4 atm of oxygen using the traveling-solvent floating-zone technique, annealed for 10 h in argon at 970 $\dg$C followed by 20 h in oxygen at 500$\dg$C\,\cite{Mang2004}. The sample was antiferromagnetic with a N\'eel temperature of about $T_N=82$ K\,\cite{Motoyama2007}.  ARPES measurements were carried out at the synchrotron radiation facility BESSY II using the UE112-PGM2a variable photon polarization  beam line and the "$1^2$"-ARPES
end station equipped with a Scienta R8000 analyzer. All measurements were performed in the normal state at $T=50$  K. The total energy resolution was set between 10 and 15 meV, while the angular resolution was $\approx 0.2\dg $. We point out that the ARPES experiments were performed at relatively high photon energies (h$\nu= 50$ to $120$ eV), a range in which the cross section for Cu $3d$ state excitations is 5 to 7 times larger than that for O $2p$ states\,\cite{Yeh1985}. The crystal was  mounted on a 6-axis cryomanipulator allowing polar, azimuthal, and tilt rotation of
the sample in ultrahigh vacuum with a precision of $0.1\dg$.  The experimental geometry is shown in Fig.~1(a). The  mirror plane $(0,0,\pi)-\Gamma-(\pi,0,0)\widehat{=}(x,0,z)$ was turned into the scattering plane. In this way cuts shown in Fig.~1(b)  parallel to the $\Gamma - (0,\pi)$ direction for various $k_x$ values  could be recorded by changing the polar angle from nearly normal incidence to more grazing incidence. In the chosen sample orientation, the Cu 3$d_{x^2-y^2}$ conduction band states have an even symmetry with respect to the scattering plane as shown in Fig.~1(a). For non-zero photoemission intensity on the mirror plane, the final state must be even with respect to this plane and the same holds for the product of the dipole operator and the initial state. Therefore, for this sample orientation and  for $p$-polarized light ($\bf{e}$  parallel to the mirror  plane, dipole operator even) the matrix element should be finite near the mirror plane, while for $s$-polarized light ($\bf{e}$  perpendicular to the mirror  plane, dipole operator odd) the matrix element should vanish (see Eq.~2). In a study of the origin of the shadow Fermi surface in cuprates it was shown that the intensity vanishes in a narrow range of $\pm 2^{\circ}$ when the transition is symmetry forbidden~\cite{Mans2006}.

ARPES intensity calculations were done fully relativistically based on the Dirac equation~\cite{Braun1996}. One-step model~\cite{Pendry1976} and multiple scattering theory have been utilized, the latter being used also for the final states. The potential for Nd$_2$CuO$_4$ has been calculated by using self consistent electronic structure calculations with the Korringa-Kohn-Rostoker method~\cite{Kaprzyk1990,Bansil1999}. In the calculations, many-body correlations were taken into account by the complex self-energy function $\Sigma$. For the initial state  we used the experimental self-energy function   derived from the ARPES experiments (see below). For the final state   a constant value $\Im\Sigma_f = 2$ eV is used.
 
In Fig.\,1(c) and (d) we present the sum of ARPES intensity plots along the edges of the BZ  recorded with right ($c_+$) and  left ($c_-$)  circularly polarized photons having two  different photon energies. For h$\nu =120$ eV [see Fig.~1(c)] and $k_x =5\pi$ [cut$\#$3 in Fig.\,1(b)] a clear HEA  at $E_H= 0.25$ eV is detected. At this energy the normal  dispersion transforms into a vertical waterfalllike dispersion accompanied by a reduction of the  intensity near $k_y=0$. On the other hand, changing the photon energy to
 h$\nu =94$ eV [see Fig.~1(d)] and $k_x=\pi$ [cut$\#2$ in Fig.~1(b)] transforms the dispersive feature into a band having a normal dispersion with a width at constant energy which increases continuously with increasing binding energy. Since we observe similar changes as a function of photon energy for $k_x=
n\pi$  $n=1, 3$, and 5 (not shown) we conclude that the change of the spectra shown in Fig.\,1(c) and (d) is caused by the variation of the photon energy but not by the change of $k_x$. Furthermore, \NCCO\ is essentially a two-dimensional electronic system\,\cite{Massidda1989} and therefore upon variation the photon energy $k_\perp$ changes, but this should not change the spectral function. Thus the drastic change of the intensity plots presented in Fig.\,1(c) and (d) signals that the HEA is  caused by matrix element effects but not by changes of the spectral function.

\begin{figure}[tb]
\centering
\includegraphics[angle=0,width=0.41\textwidth]{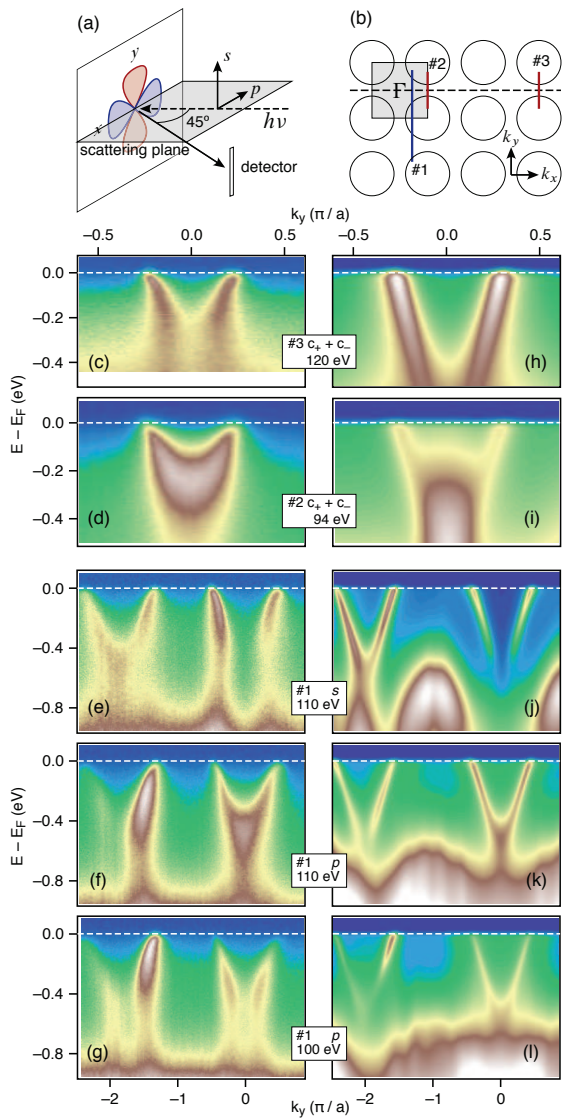}
\caption{(Color online) (a) Experimental geometry. (b) Cuts in the reciprocal space used in the present investigation. (c) Experimental ARPES intensity distribution maps  recorded with parameters given in the labels.  (h)-(l)  calculated ARPES intensities using the parameters of (c)-(g), respectively.} 
\end{figure}

In Fig.\,1 (e)-(g) we present similar energy distribution maps but now recorded  with the wave vectors  $k_x=\frac {3\pi} 8$ [cut$\#$1 in Fig.~1(b)]. In these cuts we  cross the Fermi surface at the nodal point. For h$\nu=110$ eV and $p$-polarization, near $k_y=0$ a normal dispersion is observed [see Fig.\,1(f)]. Compared to the antinodal point, the bottom of the band has moved to higher $E_B$, which is expected from band structure calculations. The spectral weight of the band extends into the  region of the non-bonding oxygen valence bands.  On the other hand,  near $k_y=-2\pi$, a HEA is observed. At the same photon energy (h$\nu=110$ eV) but for s-polarization, we know that the intensity must vanish at the mirror plane ($k_y = 0$). In Fig. 1(e) we see that, when the matrix element is intentionally suppressed in this way, a HEA appears at  $E_H = 0.4$ eV.  Thus the existence of the waterfalllike  dispersion can be unambiguously attributed  to a vanishing matrix element near $k_y=0$. In the second BZ we detect a more normal dispersion. Furthermore, for the photon energy h$\nu=100$ eV but for $p$-polarization, a HEA is observed in the first and the second BZ [see Fig.\,1(g)]. Comparing the spectra presented in Fig.\,1(f) and (g) which were measured both with the same ($p$) polarization, the difference near $k_y=0$ cannot be explained by an extinction of the matrix element due to a specific photon polarization but by an extinction of the matrix element near the $(k_x,k_y=0)$  line for specific photon energies. Comparing the intensities near $k_y=0$  presented in Fig.\,1(e) and (g), it is remarkable that the distances between the waterfalllike dispersions in momentum space along $k_y$  is in the former about twice as large as in the latter.
We attribute the differences in the energies $E_H$ (changing from 0.4 eV to 0.55 eV) to the different extinction ranges of the matrix elements.  
In Fig.\,2(a) we plot the bottom of the band along the $\Gamma-(\pi,0)$ direction, derived from fits of energy distribution curves at $k_y=0$ using spectra showing no HEA and compare this result with our density function theory (DFT) calculations. 

\begin{figure}[tb]
\centering
\includegraphics[angle=0,width=0.48\textwidth]{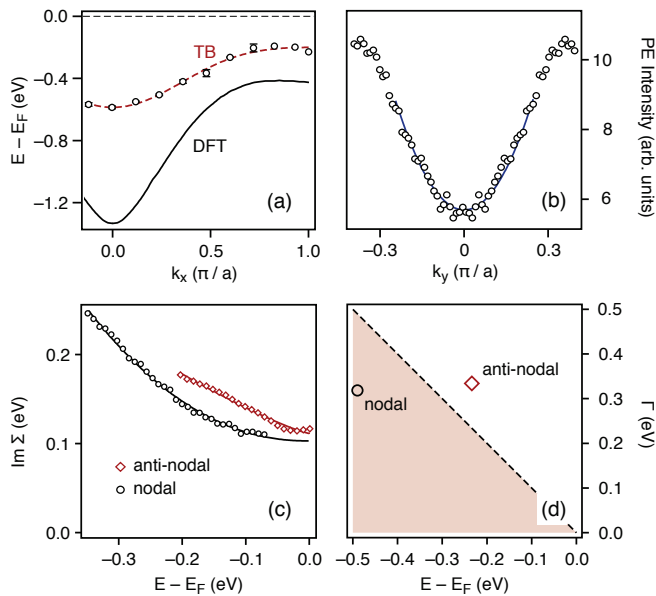}
\caption{(Color online)   (a) Bottom of the band along the (-$\pi$,0)-($\pi$,0) direction between the center of the BZ and the antinodal point. Circles: ARPES data. Dashed line: tight binding (TB) fit. Solid line: present density functional theory (DFT) calculations. (b) Momentum distribution curve of the data shown in Fig.~1(e) at $E_B=0.5$ eV, symmetrized relative to $k_y=0$. The blue line shows a quadratic momentum  dependence. (c) Imaginary part of the self-energy $\Im\Sigma$ as a function of the binding energy near the nodal and the antinodal point. The lines are fits to the data (see text).  (d)  Width $\Gamma$ of the spectral weight at the bottom of the band near the antinodal point [$k_{\|}=(\pi,0)$] and at the cut through the nodal point [$k_{\|}=(\frac {3\pi} 8,0)$]. Red region indicates the range of the existence of quasi-particles $(\Gamma <  E_B)$. } 
\end{figure}

The interpretation of the ARPES results on the HEA in terms of matrix element effects is supported by a simple calculation of the matrix element for the selected geometry and sample orientation using atomic wave functions for  the $\left|d\right>$ initial state and  $\left<p\right|/\left<f\right|$ final states\,\cite{sup}. As expected from the discussion above, these calculations yield for $s$-polarization a vanishing matrix element at the mirror plane, but for photons emitted at a finite angle relative to the mirror plane the matrix element increases linearly with that angle. This means that for small angles the intensity should increase  proportional to $k_y^2$ which is in perfect agreement with a momentum distribution curve at  $E_B=0.5$ eV of the  ARPES data shown in Fig.~1(e) [see Fig.~2(b))]. One explanation to account for the observed $h\nu $-dependence would be that for particular photon energies only final states  which are odd with respect to the mirror plane can be reached. This would explain why for $p$-polarization  the intensity is zero in the mirror plane. The calculation predicts for finite emission angles relative to the mirror plane a finite intensity  proportional to $k_y^2$  due to even final states which contribute to the matrix element  linearly with increasing angle. This would explain why for certain energies even for $p$-polarization a waterfalllike dispersion is observed [see Fig.~1(g)].

The interpretation of the HEA in terms of extrinsic matrix element effects is furthermore strongly supported by the ARPES intensity calculations which are in qualitative agreement with the  experimental ARPES data [see Fig.~1(h)-(l))]. Part of the remaining differences are related to the fact that for the non-bonding O~2$p$ bands, the same self-energy function as for the Cu~$3d$ band has been used which leads to an unphysical broadening  of the former  bands.

The presented ARPES data together with the supporting calculations provide strong evidence that the HEA is not related to  an anomalous  spectral function, i.e.,  to  specific  many-body effects. Rather, it is caused  by a wipe-out of a broad intensity distribution of the spectral weight  near particular high-symmetry lines due to the extinction of the matrix elements (see also Fig.~2 in the supplement\,\cite{sup}), i.e.,  by extrinsic effects.  We point out that  the formation of a waterfalllike  dispersion requires both a vanishing  matrix element  and a  strong broadening of the spectral weight at higher binding energies. This explains why the HEA has only been found  in strongly correlated systems in which a strong broadening of the "bands" at higher binding energies due to high scattering rates is present.

The present work can explain various results presented in the literature which were not understood previously. For example, the momentum dependence of $E_H$, when going from the nodal to the antinodal point, previously observed in \LSCO\  \cite{Chang2007a} and \NCCO\ \cite{Ikeda2009a} can be readily explained in terms of a wipe-out of the intensity of a normal quasi-particle band, in which the bottom of the band near $\Gamma$ is deeper than near the antinodal point. Furthermore, the difference of the $E_H$ values of $\approx$ 0.4  eV between $p$- and $n$-doped cuprates, which was linked to the difference of the chemical potential~\cite{Ikeda2009a} can be now understood by a wipe-out of quasi-particle bands, ranging to different energies below the Fermi level.

Since we can now follow the dispersion to the bottom of the band, we can derive reliable results for the mass enhancement  compared to DFT calculations. Using the data presented in Fig.\,2(a), we  obtain  a high-energy mass renormalization of  2.1  near the antinodal point  while at $\Gamma $  a value of 2.3 derived.  We remark that these values are connected with large errors of about 30\% since there is a considerable scattering of the energy position of the Cu-O conduction band relative to the Fermi level in the published DFT calculations. Moreover, we have  analyzed the energy-dependence of the imaginary part of the self-energy, extracted from momentum distribution curves, using the relation  $\Im\Sigma=-A-BE^\alpha$ [see Fig\,2(c)]. Near the antinodal point we obtain $A=0.111 \pm 0.003$ eV, $B= 0.47 \pm 0.09$ eV$^{1-\alpha}$ and $\alpha = 1.2\pm0.1$, while near the nodal point $A= 0.103 \pm0.002$ eV, $B= 1.4\pm0.1$ eV$^{1-\alpha}$, and $\alpha= 2.15\pm0.07$. The observation  of a nearly quadratic increase as a function of energy at the nodal point indicates a Fermi liquid behavior and explains the quadratic temperature dependence of the in-plane resistivity above the superconducting transition temperature~\cite{Tsuei1989}. The nearly  linear increase at the antinodal point  signals  the proximity to a marginal Fermi liquid~\cite{Varma2002}. Near the center of the BZ, at $k_{\|}=(\frac {3\pi} 8,0)$ the total  life-time broadening amounts to 65 $\%$ of $E_B$ [see Fig.\,2(d)]. This indicates that even at the bottom of the band, spectral weight of  quasi-particles is observed and that at these $E_B$s we are not in the incoherent range as suggested previously in the literature\,\cite{Meevasana2007,Byczuk2007,Manousakis2007,Tan2007,Zemljic2008}. In addition, the lack of a high-energy kink and the continuous increase of the  width as a function of $E_B$ do not support any scenario of a strong coupling of the charge carriers to discrete high-energy bosonic  excitations such as magnetic excitations with an energy of $2J\approx 0.3 $ eV (J is the exchange energy)~\cite{Armitage2010} which could mediate high-$T_c$ superconductivity\,\cite{Valla2007,Macridin2007,Markiewicz2007,Basak2009,Markiewicz2010}.
Rather, the results presented here indicate a strong coupling of the charge carriers to low-energy electronic excitations, e.g. spin excitations between regions near the antinodal points leading there to a marginal Fermi liquid behavior as described by~\cite{Monthoux1991,Abanov2003}. Finally, we postulate that the present results can be generalized to all cuprates since in the $p$-doped compounds very similar ARPES results, e.g. energy dependence of $E_H$, have been obtained\,\cite{Chang2007a}.

To summarize, our ARPES results on an $n$-doped cuprate together with a calculation of the ARPES intensity in a one-step model clearly show that the HEA is not related to an intrinsic anomalous dispersion of the spectral weight. Rather, it is exclusively  caused by a combination  of a wipe-out due to matrix element effects  and high scattering rates at high energies. By selecting suitable photon energies  we are able to obtain important information on the many-body properties of doped cuprates which  places strong constraints on theories of high-$T_c$ superconductivity in these systems. Furthermore, the present results provide important information about the electronic structure of doped Mott-Hubbard insulators: a vertical dispersion between the coherent quasi-particles and the incoherent lower Hubbard band\,\cite{Meevasana2007,Byczuk2007,Manousakis2007,Tan2007,Zemljic2008} is not supported.

We thank Inna Vishik for valuable comments. The work at the University of Minnesota was supported by the NSF and the NSF MRSEC program.

\bibliographystyle{phaip}
\bibliography{ARPESmany}

\end{document}


\title{Supplementary material for: The high-energy anomaly in  ARPES spectra of the cuprates---many body  or matrix element effect?}
\author{E. D. L. Rienks}
\affiliation{Helmholtz-Zentrum Berlin, Albert-Einstein-Strasse 15, D-12489 Berlin, Germany}
\author{M. \" Arr\" al\" a}
\author{M. Lindroos}
\affiliation{Department of Physics, Tampere University of Technology, PO Box 692, FIN-33101 Tampere, Finland}
\author{F. Roth}
\affiliation{Center for Free-Electron Laser Science / DESY, Notkestrasse 85, D-22607 Hamburg, Germany}
\author{W. Tabis}
\affiliation{School of Physics and Astronomy, University of Minnesota, Minneapolis, Minnesota 55455, USA}
\affiliation{University of Science and Technology, Faculty of Physics and Applied Computer Science, 30-059 Krakow, Poland}
\author{G. Yu}
\author{M. Greven}
\affiliation{School of Physics and Astronomy, University of Minnesota, Minneapolis, Minnesota 55455, USA}
\author{J. Fink}
\affiliation{Leibniz-Institute for Solid State and Materials Research Dresden, P.O.Box 270116, D-01171 Dresden, Germany}

\begin{abstract}
In this supplement, we provide  angle-dependent calculations of the matrix element for photoemission experiments described in the main paper using atomic orbitals for the Cu 3$d$ initial and the dipole allowed Cu 2$p$ or 4$f$ final states. When the matrix element is symmetry-forbidden on the mirror plane, it is found to increase linearly with increasing emission angle relative to this plane. Thus the ARPES intensity should increase quadratically with the distance $k_y$ from the mirror plane. This explains the wipe-out near the $x-z$ plane for $s$-polarization and thus the waterfalllike dispersion.
\end{abstract}

\pacs{74.25.Jb, 74.72.Ek,  79.60.i}

\maketitle

In ARPES  the measured photocurrent is given by 
\begin{equation}\label{eq:photo}
I(\omega,\boldsymbol{k})\propto |M(\omega,\boldsymbol{k})|^2 A(\omega,\boldsymbol{k})
\end{equation} 
where  $A(\omega,\boldsymbol{k})$ is the spectral function and  the matrix element 
\begin{equation}\label{eq:matrix}
 M(\omega,\boldsymbol{k}) = \left< f \right|\boldsymbol{er}\left| i \right>
\end{equation}
is determined by the final state $\left< f \right|$, the initial state $\left| i \right>$, and  the dipole operator $\bf{er}$ ($\bf{e}$ is the unit vector along the polarization direction of the photons)\,\cite{Damascelli2003}. In the following we calculate the matrix elements for the geometry and sample orientation described in the main paper using atomic initial and final states. The dipole operator for $s$- and $p$-polarization is proportional to a linear combination of $\Theta$ and $\varphi$ dependent spherical harmonics by $Y^1_1-Y^{-1}_1\propto sin(\Theta)$  and   $Y^1_1-Y^{-1}_1\propto cos(\Theta)$, respectively. $\Theta$ denotes the angle in the $z-y$ plane perpendicular to the mirror plane, while $\varphi$ is the angle around the $z$ axis.  The $\Theta$-dependence  of the matrix element is given by the relation
\begin{equation}\label{eq:photo}
M(f,i,\Theta) \propto \int \limits_{0}^\pi Y_{l_f}^{m_f}(Y^1_1\pm Y^{-1}_1) Y_{l_i}^{m_i} \sin \theta d\theta
\end{equation}
The initial state is a Cu 3$d_{x^2-y^2}$ state, hence the initial state is determined by $l=2, m=\pm 2$ and therefore $\left< i \right| \propto Y_{2}^{\pm 2}\propto sin(\theta)^2$. According to the dipole selection rules $\Delta l=\pm 1$ and $\Delta m=0,\pm 1$  excitations into $\left<p\right|$ and $\left<f\right|$ final states can occur.  In the case of $p$-polarization allowed final states are $Y_{1}^{\pm1} \propto sin(\theta)$ which is even with respect to the mirror  plane. For $p$-polarization (even dipole operator) the matrix element is finite since both the initial and the final states are even. On the other hand, for $s$-polarization (odd dipole operator) the matrix element vanishes since both the initial state and the final state are even.  In this case when  the emitted photons have a small angle $\delta$ versus the mirror plane, the final states are proportional to $sin(\Theta+\delta)\approx sin(\Theta)+\delta cos(\Theta)$. The second term now adds an odd contribution to the final state and we will have a finite matrix element which is proportional to $\delta$. Thus for small $\delta$ the intensity is $I(\delta) \propto M^2 \propto \delta^2 \propto k_y^2$ . In Fig.1 we show the calculated squared matrix element $M^2$ versus $k_y$. A similar calculation for transitions into $\left<f\right |$ final  states results in a matrix element which is also proportional to $k_y$.

\begin{figure}[tb]
\centering
\includegraphics[angle=0,width=0.45\textwidth]{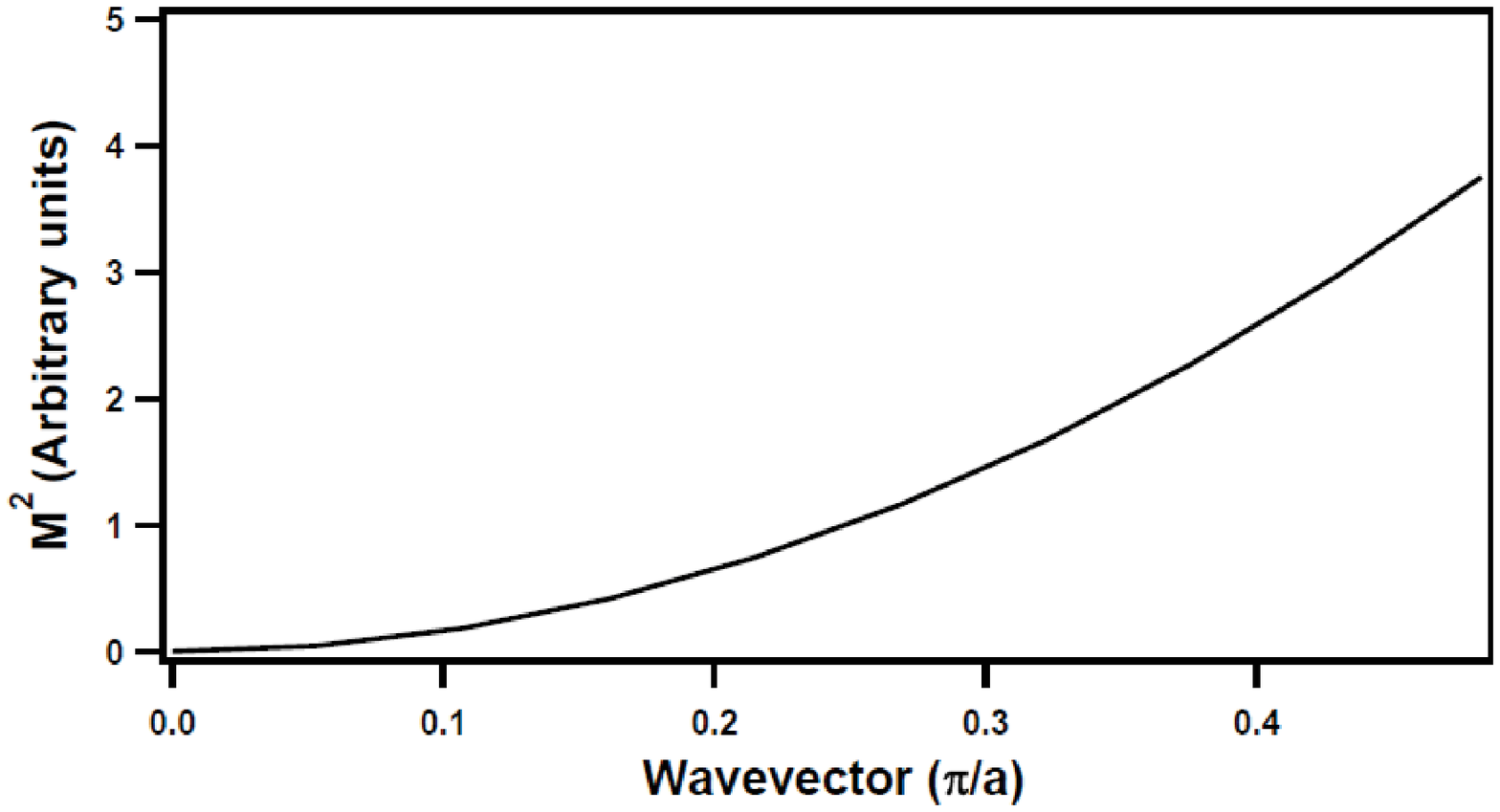}
\caption{Calculated squared matrix element $M^2$ versus the wave vector $k_y$ perpendicular to the mirror plane.} 
\end{figure}

In Fig.~2(d) of the main paper we show the intensity along $k_y$  from a momentum distribution curve at a binding energy of 0.5 eV of the ARPES intensity distribution map shown in Fig. 1(e) of the main paper. At small $k_y$ the intensity agrees perfectly with a quadratic dependence on $k_y$ indicating that the appearance of the waterfalllike dispersion can be explained in terms of a simple calculation of the matrix element using atomic initial and final state wave functions.

\begin{figure}[h!]
\centering
\includegraphics[angle=0,width=0.35\textwidth]{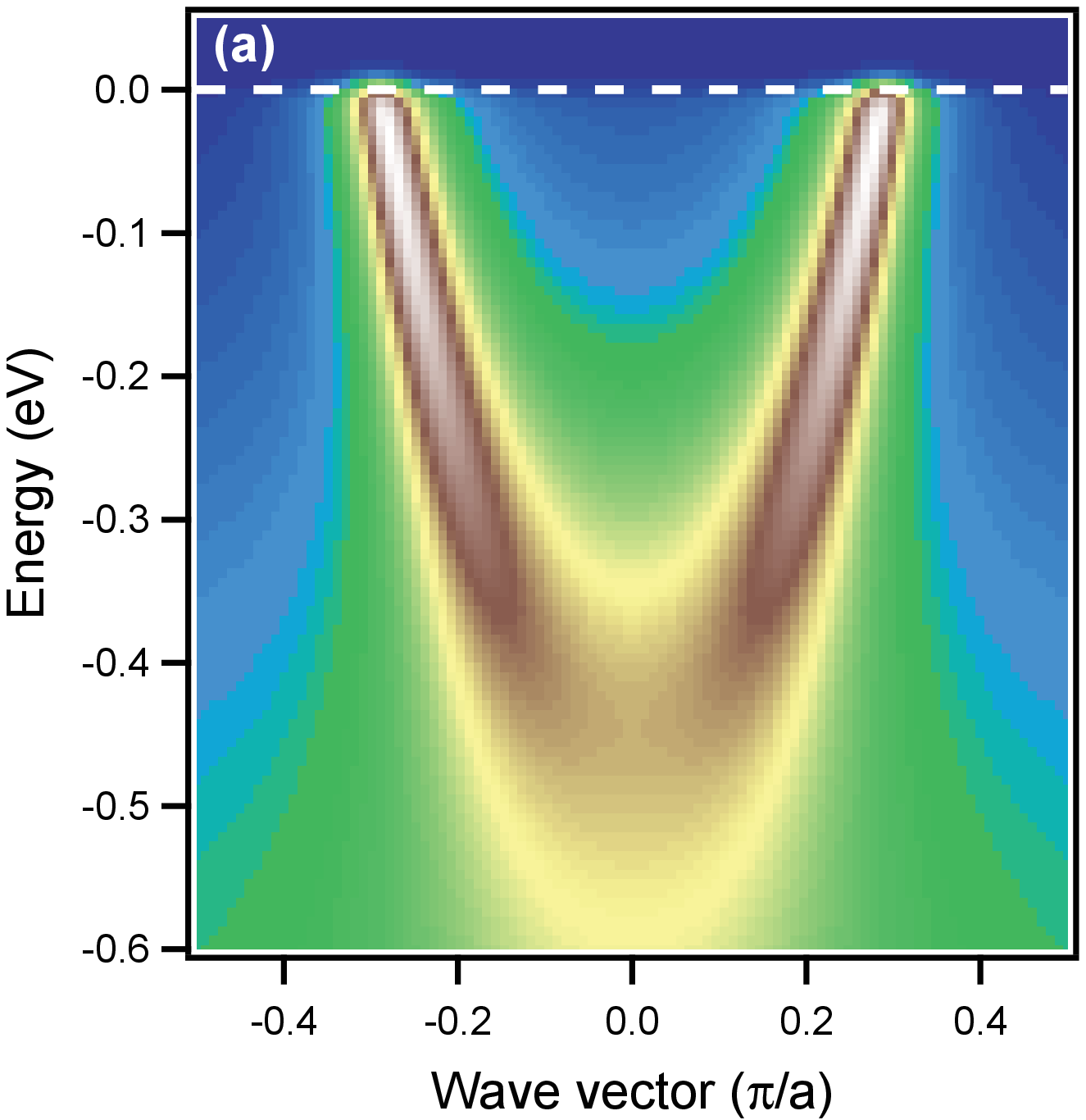}
\includegraphics[angle=0,width=0.35\textwidth]{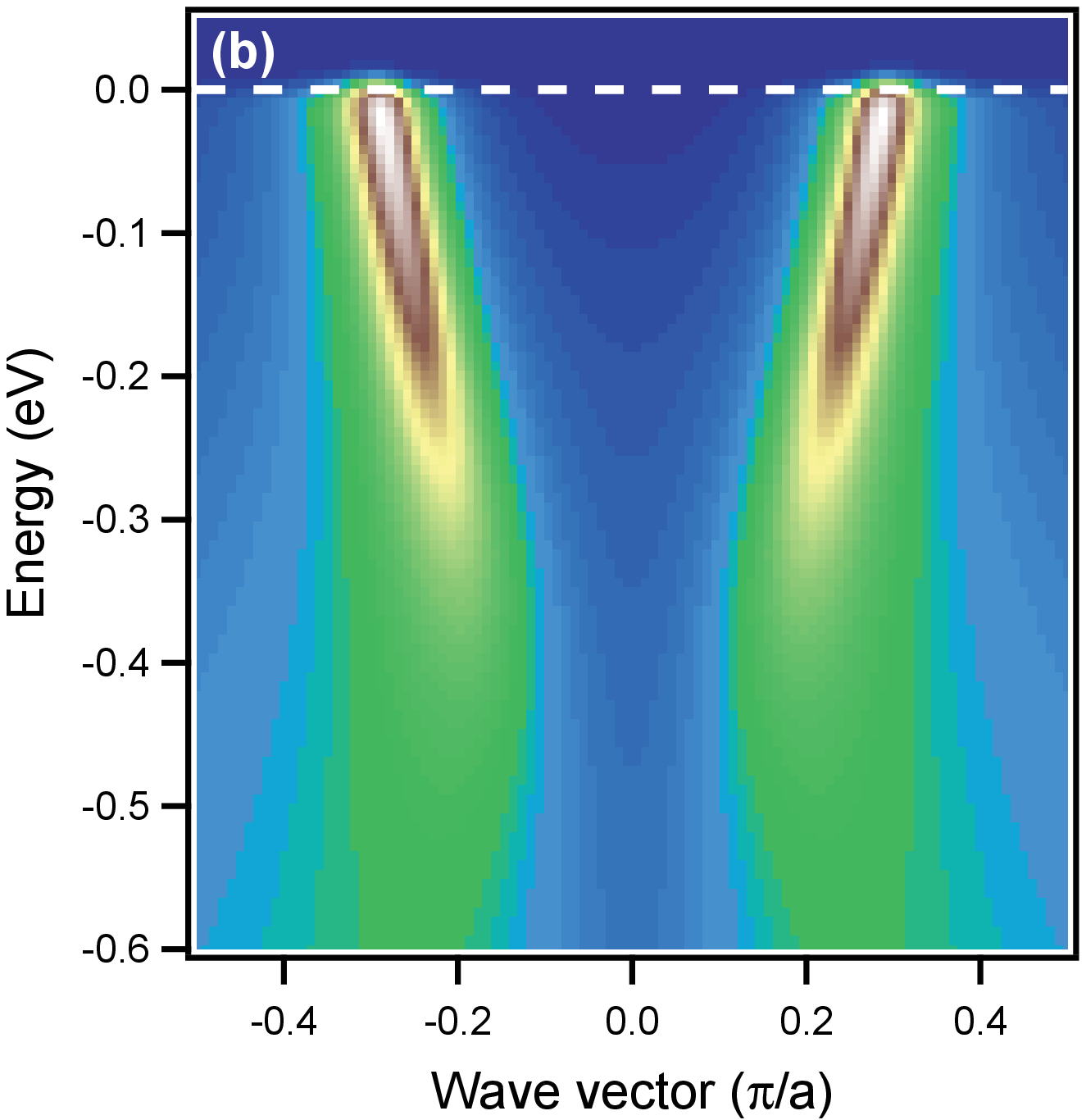}
\caption{(Color online) (a) Calculated ARPES intensity distribution map using a parabolic band together with a self-energy without a high-energy anomaly derived from the experiment. (b) The same intensity distribution map as in (a) but multiplied by a matrix element which is quadratic in the wave vector. A waterfalllike dispersion is produced.} 
\end{figure}

In Fig.~2(a) we show the calculated intensity derived from a spectral function using for convenience a parabolic band and a complex self-energy function [which shows no high-energy anomaly (HEA)] taken from the experiment at the nodal point which was mentioned in the main paper. A Shirley background~\cite{Shirley1972} was added. Multiplying the spectral function with a matrix element which is proportional to $k_y^2$ we obtain the intensity distribution map shown in Fig.~2(b) which shows a HEA. This represents a further indication that the waterfall like dispersion is caused by matrix element effects. 

Finally we discuss the energy dependence of the ARPES intensity for $p$-polarized photons using the present calculation of the matrix element [see Fig.~1(f) and (g) in the main paper]. If the final state  is even and thus proportional to $sin \Theta$ the three terms of in the integral over $\theta$ are even and thus the matrix element is finite yielding a normal dispersion without HEA. Upon changing the photon energy,  at particular energies predominantly odd final states could  be reached which are proportional to $cos \Theta$. One explanation to account for the observed $h\nu $ dependence would be that for particular photon energies only odd final states are reached. Then for $p$-polarized  photons and photoelectrons emitted in the mirror plane the matrix element is zero. Detecting electrons which are emitted at a small  angle $\delta$ relative to the mirror plane,  the final state is now  proportional to $cos(\Theta+\delta)\approx \cos \theta +\delta sin \theta$. The second term yields an even contribution to the final state proportional to $\delta$. Thus the   finite matrix element increases proportional to $\delta$ and the intensity proportional to $\delta^2$ causing a HEA similar to the case for $s$-polarized photons. In this way it is possible to explain the energy dependence of the ARPES intensity, shown in Fig.~1(f) and (g) in the main paper, by an energy dependent change of the parity of the final states.

\bibliographystyle{phaip}
\bibliography{ARPESmany}